\documentclass[a4paper,11pt]{article}
\usepackage[utf8]{inputenc}
\usepackage{amsmath}
\usepackage{amssymb}
\usepackage{graphicx}
\usepackage{xcolor}
\newcommand{\Var}{\mathrm{Var}}

\begin{document}

\title{A simple stochastic model to describe the evolution over time of core genome SNP GC content in prokaryotes}


\author{Jon Bohlin\textsuperscript{1,2,3,*}, Brittany Rose\textsuperscript{1,4}, Ola Brynildsrud\textsuperscript{1,3} \& Birgitte Freiesleben De Blasio\textsuperscript{1,4}}


\maketitle

\textsuperscript{1}Division of Infection Control and Environmental Health, Norwegian Institute of Public Health, Oslo, Norway.\par
\textsuperscript{2}Centre for Fertility and Health, Norwegian Institute of Public Health, Oslo, Norway.\par
\textsuperscript{3}Department of Production Animals, Faculty of Veterinary Medicine, Norwegian University of Life Science, Oslo, Norway.\par
\textsuperscript{4}Department of Biostatistics, Oslo Centre for Biostatistics and Epidemiology, Institute of Basic Medical Sciences, University of Oslo, Oslo, Norway.\par
\textsuperscript{*}Corresponding author\par

\begin{abstract}

Genomes in living organisms consist of the nucleotides adenine (A), guanine (G), cytosine (C) and thymine (T). All prokaryotes have genomes consisting of double-stranded DNA, where the A's and G's (purines) of one strand bind respectively to the T's and C's (pyrimidines) of the other. As such, the number of A's on one strand nearly equals the number of T's on the other, and the same is  true of one strand's G's and the other's C's. Globally, this relationship is formalized as Chargaff's first parity rule; its strandwise equivalent is Chargaff's second parity rule. Therefore, the GC content of any double-stranded DNA genome can be expressed as $\%GC=100\%-\%AT$.

Variation in prokaryotic GC content can be substantial between taxa but is generally small within microbial genomes. This variation has been found to correlate with both phylogeny and environmental factors. Since novel single-nucleotide polymorphisms (SNPs) within genomes are at least partially linked to the environment, SNP GC content can be considered a compound measure of an organism's environmental influences, lifestyle and phylogeny.

We present a mathematical model that describes how SNP GC content in microbial genomes evolves over time as a function of the AT$\rightarrow$GC and GC$\rightarrow$AT mutation rates with Gaussian white noise disturbances. The model suggests that, in non-recombining bacteria, mutations can first accumulate unnoticeably and then abruptly fluctuate out of control. Thus, minuscule variations in mutation rates can suddenly become unsustainable, ultimately driving a species to extinction if not counteracted early enough. This model, which is suited specifically to symbiotic prokaryotes, conforms to scenarios predicted by Muller's ratchet and may suggest that this is not always a gradual, degrading process. It is also in agreement with some of the empirical evidence that motivated the formulation of the Red Queen hypothesis.  We apply our model to different lineages of \textit{Renibacterium salmoninarum} and find a substantial increase in SNP GC content within the most disseminated lineage, 1a. That increase could be due to a dramatic change in environment for this lineage.

\end{abstract}

\section{Introduction}

\label{intro}
GC content varies considerably between prokaryotic species but is remarkably stable genome-wide, despite the fact that bacterial genomes are predominantly functional and expressed in some sense \cite{RochaFeil2010}. Bacteria can have an average genomic GC content of as low as 13.5\% (\emph{Candidatus} Zinderia insecticola) or of as high as 75\% (\emph{Anaeromyxobacter dehalogenans}) \cite{Bohlin2018}. While both large and small bacteria can be either GC-rich or AT-rich, there seems to be a tendency---at least in some phylogentic groups---for symbionts with smaller genomes to be more AT-rich, while soil-dwelling bacteria with large genomes tend to be more GC-rich \cite{Bohlin2014, Agashe2014}.

The mechanisms responsible for GC richness in bacteria with large genomes are poorly understood; far more can be deduced from AT-rich bacteria with small genomes (see \cite{Agashe2014} for a general review of GC content in prokaryotes). For instance, it was conjectured \cite{Bentley2004} (before being later demonstrated \cite{Hershberg2010}) that mutations are generally AT-biased due to frequent methylation of cytosine that can subsequently change to thymine. Bacteria in a symbiotic relationship with their host (often an insect) undergo reductive evolution through the loss of genes rendered unnecessary by the within-host environment. There is a clear evolutionary drive towards economizing energy expenditure \cite{McCutcheon2012, Lane2010}. When host organisms have low effective population size ($N_e$) or density, genetic drift also influences the size and base composition of symbiont genomes \cite{Wernegreen2017, Lynch2016}. The outside environment can also affect genomic base composition in bacteria \cite{Foerstner2005}.

Phylogenetic relatedness, on the other hand, exerts strong pressure against changes in GC content. This is due in large part to the significant role that protein coding genes play in bacteria and to the fact that mutations in the first two positions of a codon can change the amino acid defined by that codon \cite{Reichenberger2015}. Phylogenetic influence on base composition in prokaryotes seems to be most prominent at the genus level and below \cite{Bohlin2017}.

There are several indicators that genome size reduction occurs before genomic GC content drops \cite{Wernegreen2017}. Loss of DNA mismatch repair (MMR) genes and proofreading enzymes can nevertheless lead to a relatively quick decrease in genomic GC content \cite{AnderssonLind2008}. An increase in genomic GC content, on the other hand, can result in increased fitness \cite{Raghavan2012}, and this is associated with stronger selection on base composition \cite{Hildebrand2010, Bohlin2017, Bobay2017}. Abundance of nitrogen, as in soil, has been identified as a driver for increased genomic GC content \cite{Seward2016}.

A recent study \cite{Bohlin2018} found that single-nucleotide polymorphisms (SNPs) in microbial core genomes from different taxa were surprisingly GC-rich, except in cases where the genomes themselves were already among the most GC-rich. The study presented a mathematical model describing SNP GC content as a function of core genome GC content. The model indicated that GC$\rightarrow$AT mutations occurred at roughly double the rate of AT$\rightarrow$GC mutations, which suggests that most GC$\rightarrow$AT mutations are lost prior to fixation \cite{Bohlin2018}.

In another recent study \cite{Bohlin2019}, it was shown that while GC$\rightarrow$AT mutation rates are remarkably consistent across bacterial taxa, AT$\rightarrow$GC mutation rates vary considerably. Since the environment exerts selective pressure on bacterial base composition \cite{Foerstner2005, Reichenberger2015}, it should, at least partly, be reflected in core genome SNPs, together with evolutionary history, lifestyle and taxon.

Stochastic events strongly impact the influence of the environment on genomic base composition in bacteria. Inspired by Motoo Kimura's seminal paper \cite{Kimura1980}, we modify a previously described model \cite{Bohlin2018} to investigate SNP GC content evolution with respect to time. Furthermore, we extend the model with the assumption of Gaussian white noise perturbations in the mutation rates. We assume that SNP GC content is subject to Chargaff's parity rules. In practice, this means that core genome SNP GC content depends on the bases that are selected (including through hitchhiking \cite{Smith1974}) and not on random mutations that are purged before fixation. We employ Itô calculus to solve the stochastic differential equation (SDE) that accounts for the random perturbations in the AT$\rightarrow$GC and GC$\rightarrow$AT mutation rate parameters. We then discuss implications of the model and present outcomes that show striking concordance with Muller's ratchet \cite{Moran1996} and with evolutionary mechanisms described by the Red Queen hypothesis \cite{VanValen1973, VanValen1974}. Finally, we apply the model to a genomic data set consisting of SNP GC content differences in lineages of the fish pathogen \emph{Renibacterium salmoninarum} (taken from \cite{Brynildsrud2014}).

\section{Mathematical model}

\subsection{Motivation}

\label{Mmodel}
The mathematical model presented here is an extension of the model presented in \cite{Bohlin2018}. The original model, which describes the change in core genome SNP GC content with respect to core genome GC content, is
\begin{equation}
\label{gcMOD}
\frac{dF_{GC} (x)}{dx} = \alpha F_{GC} (x) + \beta (1-F_{GC} (x)).
\end{equation}
$x$ represents core genome GC content, while $F_{GC} (x)$ represents SNP GC content. These terms are subject to the constraints $0<x<1$ and $0<F_{GC} (x)<1$. In \cite{Bohlin2018}, the parameters $\alpha$ and $\beta$ were estimated by fitting the model to empirical data using either non-linear least square regression \cite{Bohlin2018} or Bayesian inference \cite{Bohlin2019}.

In the present study, we are concerned with the change in SNP GC content with respect to time in a stochastic setting. That is, we are now interested in the relation
\begin{equation}
F_{t+\Delta t} (\omega) = F_t (\omega) + \alpha F_t (\omega)\Delta t+\beta(1-F_t (\omega))\Delta t,
\end{equation}
where $F_t (\omega)$ represents SNP GC content at time $t$. The change in $F_t (\omega)$ with respect to trajectory $\omega$ during time $\Delta t$ is a parameter $\alpha$ times $F_t (\omega)$ times $\Delta t$ plus a parameter $\beta$ times $1-F_t(\omega)$ (SNP AT content at time $t$) times $\Delta t$. In other words, the difference in SNP GC content with respect to time is assumed to be equal to parameter multiples of SNP GC content and SNP AT content. In classical calculus notation, we write

\begin{equation}
\label{rudgcMODI}
\frac{dF_t (\omega)}{dt} = \alpha F_t (\omega) + \beta (1-F_t (\omega)),
\end{equation}
Here, $F_t (\omega)$ is a stochastic process, and we let $\alpha=a+W_t (\omega)$ and $\beta=b+W_t (\omega)$, where $a,b \in \mathbb{R}$ and $W_t (\omega)$ is a Gaussian white noise process. Equation (\ref{rudgcMODI}) is subject to the probability space $(\Omega, \mathcal{F} _t, P)$ as well as the measure space $(\mathbb{R}^+,\mathcal{G}, dt)$. $\Omega$ is the space of all trajectories $\omega$, $\mathcal{F}_t$ is its filtration with respect to each time $t\in\mathbb{R} ^+$ (\emph{i.e.} $[0,\infty)$ of which $\mathcal{G}$ is the corresponding Borel algebra and $dt$ Lebesgue measure), and $P$ is a probability measure on $\Omega$. We now have:
\begin{eqnarray}
\frac{dF_t (\omega)}{dt} &&= (a+W_t (\omega)) F_t (\omega) + (b+W_t (\omega)) (1-F_t (\omega)) \nonumber \\ \nonumber
						&&=aF_t (\omega)+F_t (\omega)W_t (\omega) +\\ \nonumber
						&&+b(1-F_t (\omega))+W_t (\omega) (1-F_t (\omega))\\ \nonumber
						&&=aF_t (\omega) + b(1-F_t (\omega))+W_t (\omega) \nonumber.
\end{eqnarray}
Hence,
\begin{equation}
\label{pregcMODI}
\frac{dF_t (\omega)}{dt} =aF_t (\omega) + b(1-F_t (\omega))+W_t (\omega).
\end{equation}
It is important to note that, in the present form, this derivative does not exist in the classical sense or in the Radon--Nikodym sense for $F_t (\omega)$. However, if we assume that $F_t (\omega)$ is a continuous semimartingale (allowing for countable and bounded jumps), the Doob--Meyer decomposition theorem (pp. 129--133 of \cite{Protter2005}) guarantees that $F_t (\omega) =F_0 + A(t) + X_t (\omega)$, where $A(t)$ is a function of bounded variation and $X_t (\omega)$ is a local martingale. Moreover, this decomposition is unique, and both $A(t)$ and $X_t (\omega)$ are adapted to $\mathcal{F}_t$. If we assume that $X_t (\omega)$ is a Brownian motion, then by chapter 3 of \cite{Oksendal2003}, (\ref{pregcMODI}) can be written as

\begin{equation}
\label{gcMODI}
dF_t (\omega)=(aF_t (\omega) + b(1-F_t (\omega)))dt+dB_t (\omega).
\end{equation}

Though the term $(aF_t (\omega) + b(1-F_t (\omega)))dt$ resembles (\ref{gcMOD}), we must handle the Brownian motion term $dB_t (\omega)$ in a non-classical way. We allow for scaled volatility $c$, as it is not unreasonable to expect variance differences across organisms and/or environments in addition to time $t$. It can be shown that a scaled Brownian motion is also a Brownian motion: Let $U_t$ be a Brownian motion (see, for instance, ch. 2 of \cite{Oksendal2003}). Then,

$$\mathbb{E}(U_t)=\frac{1}{\sqrt{2\pi t}}\int_{\mathbb{R}} ue^{-\frac{u^2}{2t}} du.$$
Letting $u=cz$ and $\frac{du}{dz}=c$, it follows that
$$\frac{1}{\sqrt{2\pi t}}\int_{\mathbb{R}} ue^{-\frac{u^2}{2t}} du=\frac{1}{\sqrt{2\pi\frac{t}{c^2}}}\int_{\mathbb{R}} ze^{-\frac{c^2z^2}{2t}}cdz$$
$$=\frac{1}{\sqrt{2\pi\frac{t}{c^2}}}\int_{\mathbb{R}} cze^{-\frac{c^2z^2}{2t}}dz=\mathbb{E}(cZ_{\frac{t}{c^2}}).$$ \par
We do not presume that $F_t (\omega)$ can see into the future. Thus, the martingale condition $\mathbb{E}(F_s (\omega)|\mathcal{F}_t)=F_t (\omega)$ with $s>t$ holds when $\mathbb{E}$ is the expectation operator with respect to probability measure $P(\omega)$, \emph{i.e.} $\mathbb{E}(X)=\int_\Omega X dP$. As such, we assume that $F_t (\omega)$ is adapted to the filtration $\mathcal{F}_t$ for each $t$, which motivates the use of the Itô integral instead of the Fisk--Stratonovich integral \cite{Protter2005}. It is therefore enough to assume that $F_t (\omega)$ is a càdlàg process, \emph{i.e.} $\lim_{s\rightarrow t^+}F_s(\omega) = F_t(\omega)$ (left-continuous with right limits; see ch. 2 of \cite{Protter2005}), implying that $F_t (\omega)$ has a countable number of bounded jumps. We can then use the Itô formula (see ch. 4 of \cite{Oksendal2003}) to solve (\ref{gcMODI}). Furthermore, since we assume that $0<F_t (\omega)<1$ and that $a$, $b$ are finite constants, it is guaranteed that (\ref{gcMODI}) has a strong and unique solution (see ch. 5 of \cite{Oksendal2003}).

First, we must identify an integrating factor that removes $F_t (\omega)$ from the right-hand side. Let
\begin{eqnarray}
dF_t (\omega) &&= (aF_t (\omega) + b(1-F_t (\omega)))dt+d\hat{B_t} (\omega) \nonumber \\
&&= (aF_t (\omega) - bF_t (\omega) +b)dt+d\hat{B_t} (\omega) \nonumber \\
&&= ((a-b)F_t (\omega) +b)dt+d\hat{B_t} (\omega), \nonumber
\end{eqnarray}
where $\hat{B_t} (\omega)$ is a $c$-scaled Brownian motion. Letting $g(t,x)=e^{(-(a-b)t)}x$, we get the integrating factor  $g(t,F_t (\omega))=Y_t (\omega)=e^{(-(a-b)t)} F_t (\omega)$. Applying Itô's formula (p. 44 of \cite{Oksendal2003}), we see that
\begin{equation}
\label{six}
dY_t (\omega) =\frac{\partial g}{\partial t} (t,F_t(\omega))dt+\frac{\partial g}{\partial t}(t, 	F_t(\omega))dF_t(\omega)+\frac{1}{2} \frac{\partial ^2g}{\partial x^2} (t,F_t (\omega))(dF_t (\omega))^2.
\end{equation}
Because $\frac{\partial ^2 g}{\partial x^2}(t,x)=0$, the last term of (\ref{six}) is equal to zero. As a result,
\begin{eqnarray}
dY_t (\omega)&&=\frac{\partial g}{\partial t} (t,F_t(\omega))dt+\frac{\partial g}{\partial x}(t, 	F_t(\omega))dF_t(\omega)\nonumber \\
&&=-(a-b)e^{(-(a-b)t)} F_t(\omega)dt+e^{(-(a-b)t)}dF_t (\omega)\nonumber \\
&&=-(a-b)e^{(-(a-b)t)} F_t(\omega)dt+ \nonumber \\
&&+e^{(-(a-b)t)}(((a-b)F_t (\omega) +b)dt+d\hat{B_t}) \nonumber \\
&&=be^{(-(a-b)t)}dt+e^{(-(a-b)t)}d\hat{B_t}. \nonumber
\end{eqnarray}

Thus, we have the differential
\begin{equation}
\label{mainDiff}
dY_t (\omega)=be^{(-(a-b)t)}dt+e^{(-(a-b)t)}d\hat{B_t},
\end{equation}
and so
$$ dY_t (\omega)=d(e^{(-(a-b)t)} F_t (\omega))=be^{(-(a-b)t)}dt+e^{(-(a-b)t)}d\hat{B_t}.$$

We can then find the formula for $F_t (\omega)$, by setting  $s\in [0,t]$, and letting
$$d(e^{(-(a-b)t)} F_t (\omega))=be^{(-(a-b)t)}dt+e^{(-(a-b)t)}d\hat{B_t}$$
which gives
$$e^{(-(a-b)t)} F_t (\omega)-F_0(\omega)=\int_0 ^t be^{(-(a-b)s)}ds + \int_0 ^t e^{(-(a-b)s)} d\hat{B_s},$$
and

\begin{equation}
\label{FinalIntegral}
F_t (\omega)=F_0(\omega) e^{(a-b)t}+\int_0 ^t be^{(a-b)(t-s)}ds + \int_0 ^t e^{(a-b)(t-s)}d\hat{B_s}.
\end{equation}
Assuming that $F_t (\omega)$ is a semimartingale, we have from the Doob--Meyer decomposition that $\int_0 ^t be^{(a-b)(t-s)}ds$ is of bounded variation and that
$$\int_0 ^t e^{(a-b)(t-s)}d\hat{B_s}$$ is a local martingale, which is in fact a martingale (see pp. 129--133 of \cite{Protter2005}). While the latter martingale term must be solved numerically, the antiderivative of the bounded variation term can be solved using the chain rule:
$$\int_0 ^t be^{(a-b)(t-s)}ds=c_0+\frac{b}{(a-b)}(e^{(a-b)t}-1).$$
We thus obtain the explicit equation for $F_t (\omega)$:
$$F_t (\omega)=F_0(\omega) e^{(a-b)t} +\frac{b}{(a-b)}(e^{(a-b)t}-1)+\int_0 ^t e^{(a-b)(t-s)}d\hat{B_s}$$
that can be written as:
\begin{equation}
\label{finalF}
F_t (\omega)=-\frac{b}{(a-b)}+(F_0(\omega) +\frac{b}{(a-b)})e^{(a-b)t}+\int_0 ^t e^{(a-b)(t-s)}d\hat{B_s}
\end{equation}
which is subject to the constraints $t\in[0,\infty)$ and $0<F_t (\omega)<1$. The integration constant $c_0$ is just assumed included in $F_0$. It should be noted that for $F_0=0$,
\begin{equation}
\label{gcMODForumla}
\mathbb{E}(F_t(\omega))=\frac{b}{(a-b)}(e^{(a-b)t}-1).
\end{equation}
Since the martingale term vanishes (see p. 30 of \cite{Oksendal2003}), we get the solution to (\ref{gcMOD}) when $t=x$ \cite{Bohlin2018}. Furthermore, we do not need to bother with the martingale term when estimating parameters $a$ and $b$. The variance is given by $\Var(F_t (\omega))=\mathbb{E}((F_t(\omega)-\mathbb{E}(F_t(\omega))^2)$, which we can solve by setting
$$A:=F_0(\omega) e^{(a-b)t}+\frac{b}{(a-b)}(e^{(a-b)t}-1)$$ and $$B:=\int_0 ^t e^{(a-b)(t-s)}d\hat{B_s}.$$ This gives:
\begin{eqnarray}
\Var(F_t (\omega))&&=\mathbb{E}((F_t(\omega)-\mathbb{E}(F_t(\omega))^2 \nonumber \\
	&&=\mathbb{E}((A+B)^2-2(A+B)A+A^2) \nonumber\\
	&&=\mathbb{E}(A^2+2AB+B^2-2A^2-2AB+A^2) \nonumber\\
	&&=\mathbb{E}(B^2)=\mathbb{E}((\int_0 ^t e^{(a-b)(t-s)}d\hat{B_s})^2). \nonumber
\end{eqnarray}
The Itô isometry (see p. 26 of \cite{Oksendal2003}) gives:
$$\mathbb{E}((\int_0 ^t e^{(a-b)(t-s)}d\hat{B_s})^2)=\mathbb{E}(\int_0 ^t (e^{(a-b)(t-s)})^2 ds)$$
$$=\int_0 ^t e^{2(a-b)(t-s)} ds.$$
We can solve $\int_0 ^t e^{2(a-b)(t-s)} ds$ explicitly by calculating its antiderivative,
$$\int_0 ^t e^{2(a-b)(t-s)} ds=d_0+\frac{1}{2(a-b)}(e^{2(a-b)t}-1).$$
Hence, we recover the expectation for $F_t(\omega)$,
\begin{equation}
\label{EgcMODI}
\mathbb{E}(F_t(\omega))=F_0(\omega) e^{(a-b)t}+\frac{b}{(a-b)}(e^{(a-b)t}-1),
\end{equation}
and the corresponding variance (integration constant $d_0$ set to zero),
\begin{equation}
\label{VargcMODI}
\Var(F_t(\omega))=\frac{1}{2(a-b)}(e^{2(a-b)t}-1).
\end{equation}

\subsection{The parameters $a$ and $b$}

\label{parameters}
We note that
\begin{equation}
\label{gcMODICond}
0<\mathbb{E}(F_t(\omega))=F_0(\omega) e^{(a-b)t}+\frac{b}{(a-b)}(e^{(a-b)t}-1)<1.
\end{equation}
For $t=0$ we see from condition (\ref{gcMODICond}) that $0<F_0(\omega)<1$. For $(a-b)>0$ $e^{(a-b) t}$ approaches inifinity so this condition is not reasonable. We are therefore left with the condition $(a-b)\leq0$. Since $0<F_0<1$ we get
$$0<F_0(\omega) e^{(a-b)t}+\frac{b}{(a-b)}(e^{(a-b) t}-1)<1$$
Letting $t\rightarrow\infty$ we see that
$$0<\frac{b}{b-a}<1$$
which implies that $b>0$ and that $a<0$.
For $a=b$ the bounded variation term $A(t)$ in eq.(\ref{finalF}) collapses into a linear equation:
\begin{equation}
  \begin{aligned}
    A(t)  &=\frac{b}{a-b}(e^{(a-b)t}-1) \\
                &=\frac{b}{a-b}(1+(a-b)t+\frac{(a-b)^2 t^2}{2!}+\cdots+\frac{(a-b)^n t^n}{n!}+\cdots-1) \\
                &=b(\frac{1}{a-b}+t+\frac{(a-b)^1 t^2}{2!}+\cdots+\frac{(a-b)^{n-1} t^n}{n!}+\cdots-\frac{1}{a-b})\\
                &=b(t+\frac{(a-b)^1 t^2}{2!}+\cdots+\frac{(a-b)^{n-1} t^n}{n!}+\cdots)\\
                &=b t
  \end{aligned}
\end{equation}
We will henceforth assume that $F_0>0$ and $(a-b)<0$.

\subsection{The martingale term}
\label{brownianMotion}
We use Gaussian white noise to model perturbations in the AT$\rightarrow$GC ($a$) and GC$\rightarrow$AT ($b$) mutation rates. We also allow for scaling of $c>0$, as mentioned above. The scale can be determined by factors such as species/strain, environment, host and presence of MMR genes. The martingale term,

\begin{equation}
\label{locMartingale}
\int_0 ^t e^{(a-b)(t-s)}d\hat{B_s},
\end{equation}
depends on the parameters $a$ and $b$ as well as on the duration of the time period. Since we assume that $(a-b)< 0$, the martingale term approaches 0 as $t\rightarrow\infty$ and Brownian motion $\hat{B_t} (\omega)$ for $a=b$. For $(a-b)<0$ it can be seen that (\ref{locMartingale}) increases as $s\rightarrow t$.

We can reach the same conclusion by examining the variance of $F_t (\omega)$ (described in (\ref{VargcMODI}) above). The Brownian motion is assumed to have mean $\mu=0$ and variance $\mathbb{E}(\hat{B^2_t}(\omega) )=t$. Thus, the variance of Browninan motion is in general expected to increase with time $t$. Since there is no simple way to calculate the integral in (\ref{locMartingale}) analytically, we do so numerically:
\begin{equation}
\label{numLocMartingale}
\int_0 ^t e^{(a-b)(t-s)}d\hat{B_s}=\sum_{s_0} ^{s_N} e^{(a-b)(t-s_i)}(\hat{W}_{s_{i+1}} (\omega) - \hat{W}_{s_i}(\omega))\Delta s_i,
\end{equation}
where $\hat{W_s} (\omega)$ is $c$-scaled white noise, $\Delta s_i =s_{i+1}-s_i$, and $s_0=0,\ldots,s_i=t_i,\ldots,s_N=t$.

\subsection{The Girsanov transform}

Equation (\ref{mainDiff}) can be written as
$$dF_t (\omega)=((a-b)F_t (\omega) +b)dt+d\hat{B_t} (\omega).$$
Since we know from (\ref{finalF}) that
$$F_t (\omega)=-\frac{b}{(a-b)}+(F_0(\omega) +\frac{b}{(a-b)})e^{(a-b)t}+\int_0 ^t e^{(a-b)(t-s)}d\hat{B_s}.$$ If we let
$$Z_t (\omega)= \int_0 ^t e^{(a-b)(t-s)}d\hat{B_s},$$
then $Z_t (\omega)$ is an Itô process (see \cite{Oksendal2003}). After some rearrangements we can set
$$K_t (\omega) = (a-b)((F_0(\omega)+\frac{b}{a-b}) e^{(a-b)t} +Z_t (\omega)),$$
and since $Z_t (\omega)$ is a martingale, we know from the Doob--Meyer decomposition that $K_t (\omega)$ is also a martingale. We can thus write
$$dF_t(\omega) = K_t(\omega)dt+d\hat{B}_t (\omega).$$
The Girsanov theorem allows us to compute the Radon--Nikodym derivative (see ch. 3, p. 146 of \cite{Protter2005}) of a measure $Q$ with respect to the probability measure $P$ as follows:

$$\frac{dQ}{dP}=exp{(-\int_0 ^t K_s(\omega) d\hat{B}_s- \frac{1}{2}\int_0 ^t K_s ^2(\omega) ds)}.$$
This means that $F_t (\omega)$ is a Brownian motion under the measure $Q$, since we assume that $(a-b)< 0$ which implies that Kazamaki's (and hence Novikov's condition) apply $\forall t$ (see chs. 4 and 8 of \cite{Oksendal2003}).

\subsection{Further generalizations}

The model describing SNP GC content can be made more general if we assume that the parameters $a$ and $b$ are functions. It is important to note that if $a$ and $b$ are functions with respect to time, obtaining an analytical solution may be impossible. While up to this point we have assumed that variation in the model is described by a white noise process, a more complicated noise term $X_t$ could also be used. For instance, if we let
$$\frac{dF_t (\omega)}{dt} = (a+X_t (\omega)) F_t (\omega) + (b+X_t (\omega)) (1-F_t (\omega)),$$
we have
$$\frac{dF_t (\omega)}{dt}=aF_t (\omega) + X_t (\omega)F_t (\omega) + b+X_t (\omega) - (b+X_t (\omega))F_t (\omega).$$
This reduces to
$$\frac{dF_t (\omega)}{dt}=(a-b)F_t (\omega) + b+X_t (\omega),$$
where $$X_t (\omega) = \theta(t, \omega)+\kappa (t, \omega) \hat{W_t} (\omega).$$
Thus,
$$\frac{dF_t (\omega)}{dt}=aF_t (\omega) + b(1-F_t (\omega))+(\theta(t, \omega)+\kappa (t, \omega) \hat{W_t} (\omega)),$$
and after rearranging:
\begin{equation}
\label{generalgcMOD}
dF_t (\omega)=((a-b)F_t (\omega) +\theta(t, \omega)+b)dt+\kappa (t, \omega) dB_t (\omega).
\end{equation}

We could, for instance, let $X_t (\omega)$ be a mean-reverting Ornstein--Uhlenbeck process, \emph{i.e.}
$$\frac{dX_t (\omega)}{dt}=GC_0-F_t (\omega)+\hat{W_t}(\omega).$$
Hence, we let $\theta(t, \omega)=GC_0-F_t (\omega)$ and $\kappa(t, \omega)=1$. Plugging these into (\ref{generalgcMOD}), we see
\begin{equation}
\label{OUgcMOD}
dF_t (\omega)=((a-b)F_t (\omega) +GC_0+b)dt+d\hat{B_t} (\omega).
\end{equation}
We can now use the integrating factor $g(t,F_t (\omega))=Y_t (\omega)=e^{(-(a-b)t)} F_t (\omega)$ to solve (\ref{OUgcMOD}) in a similar fashion to (\ref{FinalIntegral}).

\section{Results and Discussion}

\begin{figure}
\includegraphics[width=0.7\textwidth,angle=-90]{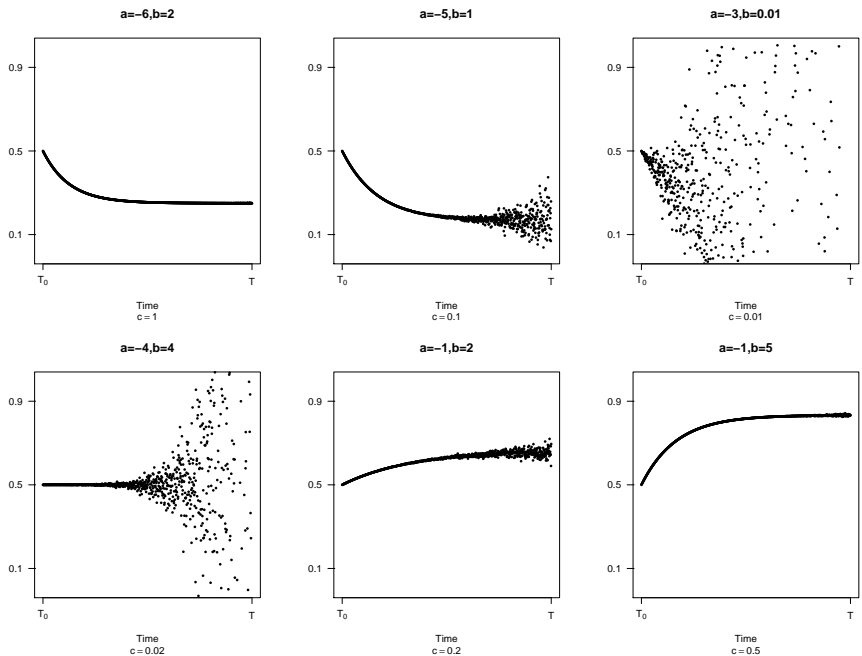}
\caption{The model (\ref{finalF}) with different combinations of parameters $a$, $b$ and Brownian scaling coefficient $c$, all starting at $F_0=0.5$. The vertical axis describes SNP GC content, while the horizontal axis describes time $t$ from $T_0=0$ to $T=1$.}
\label{fig:1}
\end{figure}

Equation (\ref{finalF}) describes a model for core genome SNP GC content in prokaryotes. It obeys Chargaff's parity laws \cite{Chargaff1954}. As discussed in section \ref{intro}, SNPs are subject to natural selection, which is in turn mediated by the environment of the organism(s) at hand. While a drop in SNP GC content could indicate relaxed selective pressures with ensuing mutations from genetic drift and AT mutational bias \cite{Hershberg2010}, increased negative or purifying selection may favor GC-biased SNPs \cite{RochaFeil2010, Hildebrand2010, Bohlin2017, Bobay2017}. Selective pressure for improved fitness could also lead to increased GC content \cite{Raghavan2012}. Carbon starvation \cite{Hellweger2018} and/or nitrogen abundance \cite{Seward2016} have also been found to have an effect on genomic base composition.

All in all, microbial organisms in the same environments often acquire the same nucleotide biases if enough time is allowed to pass \cite{Foerstner2005, Reichenberger2015}. Such environmental signatures become particularly evident in SNPs since, as discussed above and in section \ref{intro}, these polymorphisms arise as a consequence of natural selection regulated by the environment. By fitting (\ref{finalF}) to empirical data with either non-linear regression models or Bayesian inference, we can estimate the relative proportions of mutation from AT$\rightarrow$GC ($a$) and GC$\rightarrow$AT ($b$) over time. The model described by (\ref{gcMOD}) estimates analogous parameters for AT$\rightarrow$GC ($\alpha$) and GC$\rightarrow$AT ($\beta$), but with respect to core genome GC content rather than time.

\subsection{The accumulating effects of stochastic processes}
Figure \ref{fig:1} shows different paths of SNP GC content with respect to time; increase, decrease and stasis for various Brownian scaling coefficients $c$.

All stochastic fluctuations observed in the curves in figure \ref{fig:1} are a consequence of the Brownian motion term (\ref{locMartingale}). These SNP GC content curves all become more unstable as time passes, to varying degrees depending on $c$. The mathematical mechanisms behind these stochastic fluctuations are outlined in section \ref{brownianMotion}. Equation (\ref{locMartingale}) indicates that both the mutation parameters $a$ and $b$ are responsible for how the stochastic fluctuations progress with respect to time (see also section \ref{parameters}).

Some of the paths in figure \ref{fig:1} initially exhibit barely visible stochastic fluctuations, but these grow in magnitude as the SNP GC content mutation rates start to vary out of control, especially for low values of $c$.

The progression of the stochastic fluctuations in the SNP GC content curves for low $c$ is not at all expected \emph{a priori}. Below, we demonstrate that the nature of the abruptly exploding mutation rates is supported both in theory \cite{Moran1996, VanValen1973, VanValen1974} and in practice. In the following examples, we focus on mechanisms resulting in genome reduction \cite{McCutcheon2012, Wernegreen2017} and subsequent AT bias in the base composition of many symbionts. We end this section with a case study utilizing real genetic data on SNP GC content in different lineages of the fish pathogen \emph{R. salmoninarum}.

\subsection{Evolution of microbial obligate symbionts}

\label{symbionts}
Free-living bacteria that develop a sustained symbiotic relationship with a host \cite{Boscaro2017} will often, over time, undergo genome reduction \cite{McCutcheon2012}. This process of genome reduction is preceded by a phase of pseudogenization, in which the symbiont's genome retains its usual size \cite{Klasson2017}, but the genes are not under selective constraints imposed by the host and thus become abundant. Accumulated mutations eventually render many genes defective \cite{Wernegreen2017,Moran2014}. The process of pseudogenization may drag on for a long time \cite{Hosokawa2016}, but eventually non-expressed genetic regions will be excised due to energy economization \cite{Lane2010}, lack of recombination, and/or the absence of streamlining due to low population density and reduced selective pressure from the environment \cite{Lynch2016,Moran1996}. The first genes lost are typically those least conserved within a species \cite{Hershberg2016}. It is only after a continuous symbiotic relationship and the pseudogenization phase that a drop in genomic GC content seems to occur, most likely because of the loss of MMR genes that counter the AT mutational bias  \cite{McCutcheon2012}.

After the decrease in genomic GC content, there is usually no return to a free-living lifestyle for the bacterium \cite{Hosokawa2016}. Non-recombining symbionts ultimately disintegrate, as described by the concept of Muller's ratchet \cite{Moran1996}. According to some recent findings, the host, which eventually becomes dependent on the symbiont, can establish similar relationships with other bacteria \cite{Moran2014, Wernegreen2015, Hosokawa2016, Wernegreen2017, Boscaro2017}.

Intracellular pathogens, on the other hand, do not appear to engage in symbiotic relationships with a host, most likely due to the increased constraints of a pathogen-host relationship \cite{Weinert2017}. As such, though these pathogens may undergo genome reduction, they do not seem to experience the same dramatic gene loss observed in some symbionts, which are reduced to mere organelles in their hosts \cite{Moran2014}. It is not uncommon, however, for the genomic base composition of intracellular pathogens to be AT-biased \cite{Weinert2017}.

There do appear to be some similarities between the evolutionary mechanisms of symbionts and those of free-living bacteria that undergo changes in environment even if not through attachment to a host. There are only a few documented examples of free-living bacteria that experience genome reduction with subsequent genomic AT bias after a change in environment/niche. One of these is the cyanobacterium \emph{Prochlorococcus spp.} \cite{Martinez2015}, whose high-light ecotypes living close to the water surface are more AT-rich and have smaller genomes than the low-light ecotypes living at greater depths \cite{Batut2014}. Indeed, genomic GC content and genome size increase, respectively, from 30.8\% and 1.66 megabase pairs (Mbp) in the high-light strains to 50.0\% and 2.68 Mbp in the low-light strains \cite{Batut2014}.

\subsection{Modeling AT bias in microbial genomes}

As mentioned above, microbial genomes appear to become more AT-rich after the loss of MMR genes, regardless of niche and/or environment. This is most likely due to AT mutational bias \cite{Hershberg2010}, and it may be mediated by genetic drift in light of relaxed selective pressures \cite{McCutcheon2012}.

The model in (\ref{locMartingale}) was formulated in a recent study \cite{Bohlin2018} that assumed that the change in SNP GC content with respect to core genome GC content was a constant multiple of SNP GC content and another constant multiplied by SNP AT content. In the present study, we investigated how SNP GC content evolves over time, allowing for stochastic fluctuations. We modeled these fluctuations using a Gaussian white noise process $\hat{W_t} (\omega)$, which is subject to a scaled $c>0$, in the AT$\rightarrow$GC and GC$\rightarrow$AT mutation rates. We introduced the scaling to account for differences between species, environment/niche, and selective pressures or lack thereof.

From figure \ref{fig:1}, it can be seen that the mutation rates remain fairly stable, at least in the increasing and decreasing curves, before abruptly fluctuating out of control. Once we decrease $c$ the random fluctuations start sooner and escalate a bit more. Since the mutation rates fluctuate so drastically as time $t\rightarrow T$, it is natural to expect that the outcome predicted by Muller's ratchet will be achieved \cite{Moran1996}, \emph{i.e.} that the bacterial population will go extinct. However, (\ref{finalF}) suggests that although the random fluctuations start relatively late, the species' fate may be sealed far earlier, before any stochastic fluctuation can be observed.

A loss of MMR genes could imply that the scaling parameter $c$ adds more weight to the martingale term (\ref{Mmodel}), which triggers the amplification of the stochastic fluctuations. However, the similarity of the mutation rate parameters $a$ and $b$ can also influence the magnitude of the stochastic fluctuations. Indeed, from (\ref{locMartingale}), it can be seen that low mutation rates magnify the effect of the martingale term as $a-b\rightarrow 0$, since $e^{(a-b)}\rightarrow 1$.

\subsection{Connections with theories from evolutionary biology}

Leigh Van Valen wanted a model to confirm that extinction rates correlate with age in the fossil record. However, after testing this hypothesis, he found no such correlation \cite{VanValen1973}. Thus, he formulated the Red Queen hypothesis, taking its name from Lewis Carroll's 1871 book \emph{Through the Looking-Glass, and What Alice Found There}. In that book, the Red Queen utters to Alice about the nature of Looking-Glass Land, "Now, here, you see, it takes all the running you can do, to keep in the same place."

Later on, the Red Queen hypothesis was expanded to account for molecular data as well \cite{VanValen1974}. The model presented in (\ref{finalF}) demonstrates related mechanisms for prokaryotes and sheds light on the case of microbial symbionts that have undergone genome reduction with a subsequent drop in GC content. If mutations are not kept in check, extinction will ensue. In other words, the martingale term (\ref{locMartingale}) must be kept as low as possible in order to avoid the random fluctuations that lead to extinction. Since the choice of $c$ is shaped by factors such as species, environment, host and mutation rates, extinction rates will differ between populations, as predicted by the Red Queen hypothesis (See also Figure \ref{fig:1}). Furthermore, non-recombining clonal organisms will sooner or later accumulate deleterious mutations that decrease the organisms' fitness to the point of driving their species to extinction. 

In the previous section, we discussed how microbial symbionts undergo genome reduction together with a drop in GC content, most likely as a consequence of lost MMR genes. The genomes of these symbionts eventually disintegrate due to accumulated hitchhiking effects \cite{Smith1974} and genetic drift, as posited by Muller's ratchet \cite{Moran1996}. There are experimental findings to support these hypotheses \cite{Lenski2011}. Our model in (\ref{finalF}) provides insight to this by delineating the stochastic fluctuation in mutation rates that will ultimately spiral out of control, depending on the mutation parameters $a$ and $b$ and on the scaling parameter $c$.

\subsection{Mutation rates in \emph{R. salmoninarum}}

\begin{figure}
\includegraphics[width=0.6\textwidth,angle=-90]{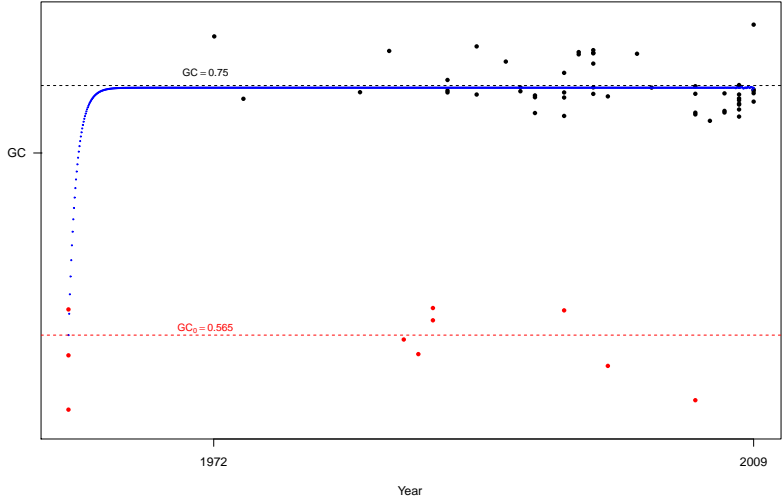}
\caption{\emph{R. salmoninarum} lineage 1a SNP GC content (vertical axis) plotted against year (horizontal axis). SNP GC content of lineages 1b and 2 is similar to \emph{R. salmoninarum} genomic GC content (red line).}
\label{fig:2}
\end{figure}

The fish pathogen \emph{R. salmoninarum} is the causative agent of bacterial kidney disease (BKD), which predominantly afflicts salmonoids. \emph{R. salmoninarum} belongs to the GC-rich, gram-positive Actinobacteria family. It is an intracellular pathogen with a genome size of approximately 3.15 Mbp and a genomic GC content of 56.5\%. Its genome is remarkably well conserved and thus appears not to recombine \cite{Brynildsrud2014}.

In a previous publication, we examined the SNP GC content over time of \emph{R. salmoninarum} lineage 1, consisting of sublineages 1a (isolated from North America and Europe) and 1b (isolated from North America), and of lineage 2 (isolated from the UK and Norway) \cite{Brynildsrud2014}. While sublineage 1b (3 isolates) and lineage 2 (7 isolates) are more endemic to particular environments, sublineage 1a (52 isolates) is widely dispersed across North America and Norway.  We found that SNP GC content in sublineage 1b and in lineage 2 was equal to 56.5\%, \emph{i.e.} the genomic GC content. The SNP GC content of sublineage 1a, however, was approximately 75\% (see figure \ref{fig:2}).

Therefore, we here set $c=1$ and $F_0 =0.565$ to correspond to the genomic GC content. We estimate $a$ and $b$ using Bayesian inference, selecting non-informative uniform distributions as priors for both parameters. The median posterior estimates are $a=-22.668$ and $b=67.421$, which suggests that the martingale term (\ref{locMartingale}) holds minimal influence. Furthermore, the GC$\rightarrow$AT mutation rate is substantially higher than the AT$\rightarrow$GC mutation rate; there is a ratio of almost 3:1 between them ($67.421/22.668=2.98$), which is highly unusual \cite{Bohlin2019}.

Though the increased SNP GC content is puzzling, it may indicate that sublineage 1a is subject to stronger selective pressure than sublineage 1b and lineage 2 \cite{Hildebrand2010, Bobay2017}, since recombination is not known to take place in \emph{R. salmoninarum}. Recent publications argue that nitrogen abundance and carbon starving, which can occur at great ocean depths, may push for increased GC content \cite{Seward2016,Hellweger2018}, but further research is needed before any conclusion can be drawn.

\section{Conclusions}

We have presented a mathematical model that describes change in SNP GC content over time as a function of mutation parameters $a$ and $b$. The model contains a stochastic term that describes how minuscule, random changes in mutation rates early on can lead to abrupt, disastrous fluctuations later. We treated examples of this phenomenon in host-associated and symbiotic bacteria.

The model, with its incorporated stochastic term and corresponding scaling parameter $c$, shows remarkable congruence with at least some parts of the Red Queen hypothesis. In the model, $c$ must not be too large to avoid genomic disintegration. Varying $c$ among species implies that the lifespan of a species need not correlate with the time to its extinction. Furthermore, the model demonstrates how Muller's ratchet operates and suggests that extinction may occur rapidly, depending on $c$, as opposed to arising slowly.

When we applied the model to different lineages of \emph{R. salmoninarum}, we found that one sublineage, 1a, exhibited a dramatic increase of approximately 20\% in SNP GC content, while no differences were detected in sublineage 1b or lineage 2. Dramatic drops in genomic GC content have been documented in both host-associated and free-living bacteria; increases in genomic GC content are less common. The substantial rise in SNP GC content observed in sublineage 1a may thus be the start of a process leading to increased genomic GC content. Since recombination is absent, or at least very rare, in \emph{R. salmoninarum}, the increase in SNP GC content may also indicate that the genome is subject to increased selective pressure, which drives the genomic GC content upwards. Alternatively, sublineage 1a could have moved to a different environmental niche than sublineage 1b and lineage 2, one in which carbon is scarce \cite{Hellweger2018}.

Our use of stochastic differential equations, which allow for deterministic modeling of random processes, revealed how bacterial mutation rates may be influenced by stochastic fluctuations. Although simple, the model described here provides novel insight into evolutionary processes with mathematical rigour.

\section{Materials and Methods}

The genomes utilized in our study were taken from a previous publication \cite{Brynildsrud2014}. They are all available from the European Bioinformatics Institute (accession number: PRJEB4487). The genomic data files were assembled using MAQ 0.7.1 \cite{Li2008} against the reference \emph{R. salmoninarum} ATCC33209 (NCBI accession number: NC 010168.1), as described in \cite{Brynildsrud2014}.

In the present study, 6 isolates were excluded due to missing date information or poor assembly quality. The removed isolates were Rs3, 5223, 684, MT3106, Cow-chs-94 and NCIMB 1111 (see \cite{Brynildsrud2014} for details). SNPs were extracted using parSNP from HarvestTools \cite{Treangen2014}, and Seaview \cite{Seaview} was used to examine the base composition of the SNPs to confirm that Chargaff's parity laws were followed, \emph{i.e.} to verify that there were approximately similar numbers of A's and T's and of G's and C's. Both sublineage 1a (52 isolates, $>1400$ SNPs) and sublineage 1b (3 isolates, $>400$ SNPs) conformed to these rules within 2\%, while a 5\% deviation was found for lineage 2 (7 isolates, $>500$ SNPs). This deviation could be due to recent mutations, natural selection and/or sequencing/assembly errors, as SNP base composition was similar to genomic GC content (\emph{i.e.} 56.5\%).

All figures were generated and statistical analyses performed in R \cite{R2018}. Bayesian parameter estimates were obtained using JAGS \cite{JAGS}. Non-informative uniform priors from --100 to 100 were assumed for both $a$ and $b$, and model precision was set to 1.0E--2. The Markov chain ran for 5,000,000 iterations with thinning set to 1,000. 12,500 iterations were saved. All chains converged.

\end{document}